\documentclass[aps,twocolumn,prl,amsmath,amssymb,showpacs,superscriptaddress]{revtex4}
\usepackage{multirow}
\usepackage{graphicx,epsfig}
\usepackage{wasysym}
\usepackage{color,soul} 
\usepackage{chngcntr}

\def \beq {\begin{equation}}
\def \edq {\end{equation}}
\def \bes {\begin{subequations}}
\def \eds {\end{subequations}}
\def \beqn {\begin{equation*}}
\def \edqn {\end{equation*}}

\def \dag {\dagger}

\def \veps {\varepsilon}

\def \calh {{\cal{H}}}

\def \calg {{\cal{G}}}

\def \hPsi {\hat{\Psi}}

\def \ha {\hat{a}}
\def \hb {\hat{b}}

\newcommand{\be}{\begin{equation}}
\newcommand{\ee}{  \end{equation}}
\newcommand{\ba}{\begin{eqnarray}}
\newcommand{\ea}{  \end{eqnarray}}

\let\Re\undefined

\DeclareMathOperator{\Re}{Re}

\begin{document}
\title{Dynamical energy transfer in ac driven quantum systems}
\author{Mar\'{\i}a Florencia Ludovico}
\thanks{These authors contributed equally to this work.}
\affiliation{Departamento de F\'{\i}sica, FCEyN, Universidad de Buenos Aires and IFIBA, Pabell\'on I, Ciudad Universitaria, 1428 CABA Argentina}
\author{Jong Soo Lim$^{\ast}$}
\affiliation{Instituto de F\'{\i}sica Interdisciplinar y Sistemas Complejos
IFISC (UIB-CSIC), E-07122 Palma de Mallorca, Spain}
\affiliation{School of Physics, Korea Institute for Advanced Study, Seoul 130-722, Korea}
\author{Michael Moskalets}
\affiliation{Department of Metal and Semiconductor Physics,
NTU "Kharkiv Polytechnic Institute", 61002 Kharkiv, Ukraine}
\author{Liliana Arrachea}
\affiliation{Departamento de F\'{\i}sica, FCEyN, Universidad de Buenos Aires and IFIBA, Pabell\'on I, Ciudad Universitaria, 1428 CABA Argentina}
\author{David S\'anchez}
\affiliation{Instituto de F\'{\i}sica Interdisciplinar y Sistemas Complejos
IFISC (UIB-CSIC), E-07122 Palma de Mallorca, Spain}
\affiliation{Kavli Institute for Theoretical Physics, University of California,
Santa Barbara, California 93106, USA}

\begin{abstract}
We analyze the time-dependent energy and heat flows in a resonant
level coupled to a fermionic continuum. The level
is periodically forced with an external power source
that supplies energy into the system. Based
on the tunneling Hamiltonian approach and scattering theory,
we discuss the different contributions to the total energy flux.
We then derive the appropriate expression for the dynamical dissipation,
in accordance with the fundamental principles of thermodynamics.
Remarkably, we find that the dissipated heat can be expressed
as a Joule law with a universal resistance that is constant at all times.
\end{abstract}


\maketitle

Quite generally, energy
flows through a physical system coupled to
a power source. In the last decades, typical system sizes
have been reduced to the nanoscale and, as a consequence,
energy transfer is to be treated quantum-mechanically~\cite{Jezouin:2013fx}.
Fundamental aspects of light-powered biological energy transport~\cite{lam13},
thermoelectric waste heat recovery~\cite{hic93},
and ultimate refrigeration protocols~\cite{gia06}
have been recently uncovered using quantum mechanical principles.
However, most discussions are limited
to stationary or time-averaged properties~\cite{Giazotto:2012cs,Lee:2013fc,Sothmann:2013ch,Koski:2013il,Brantut:2013wp}.

Time-dependent quantum transport reveals the dynamical
scales that dominate charge transfer across
phase-coherent conductors \cite{but93a,but93b}.
A prominent example is the experimentally realized
quantum capacitor, which exhibits a pure ac reponse~\cite{gab06,feb07}.
Applied time-periodic potentials also become a crucial
tool to generate directed transport of charge and spin in spatially
asymmetric ratchet-like systems~\cite{lin99,cos10}
and to control matter tunneling in Bose-Einstein condensates~\cite{lig07}.
Furthermore, the study of ac-driven quantum systems
sheds light on the role of fluctuating forces
in nanoelectromechanical resonators~\cite{ste09,las09}.
Several aspects related to time-dependent energy transport in electron systems
have been also investigated. Heat production in nanoscale
engines is discussed in Refs.~\cite{arr07,mos09}
while molecular heat pumping against thermal gradients
is proposed in Ref.~\cite{seg06}. Furthermore,
the concept of local temperature in ac pumps
has been generalized in Ref.~\cite{cas11}
whereas universal thermal resistance has been
predicted for low-temperature dynamical transport in Ref.~\cite{lim13}.
\begin{figure}[!h]
\begin{center}
\includegraphics[width=0.45\textwidth]{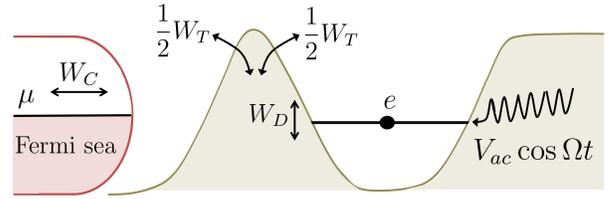}
\caption{Energy diagram of the system under consideration. A single electronic level
(the impurity with charge $e$) is coupled to a Fermi sea (the reservoir
with chemical potential $\mu$).
Energy is supplied into the system by a power source (amplitude $V_{ac}$
and frequency $\Omega$) attached to the quantum level.
Thus, energy rates are created not only at the impurity ($W_D$)
but also at the reservoir ($W_C$) and in the contact region ($W_T$).}
\label{fig1}
\end{center} 
\end{figure}

Here, we aim at the time-resolved energy production and redistribution
in ac-driven quantum coherent electron systems.
We show that the coupling between the different parts of the system not only provides a necessary mechanism for particle exchange, like in the case of charge transport, but also contributes to the energy transport. This contribution is of ac nature. Though the time-average of this energy vanishes, it allows for a temporary energy storage. Therefore, the coupling region can be referred to as an {\em energy reactance}, which only affects peak power developed in the dynamics. Our goal is also to discuss which portion of the time-resolved energy can be identified as heat, in accordance to the fundamental laws of the thermodynamics.

To be more precise let us consider a simple but generic model,
the resonant level model sketched in Fig.~\ref{fig1}.
It describes a localized fermion (the impurity) coupled to a
fermionic band of continuous density of states (the reservoir).
This model has been widely used across disciplines to study
asymmetric atomic spectra~\cite{fan61},
dissipative quantum mechanics~\cite{gui85}
and resonant-tunneling semiconductor heterostructures~\cite{jau94}, to name a few.
Transitions from the quantum level to the reservoir
yield a finite lifetime to the localized fermion
which can be represented with a Lorentzian density of states.
We consider the case in which the level is attached to
a harmonically driven power source as in Fig.~\ref{fig1}. Then,
the Hamiltonian reads,
\beq\label{eq_H}
\calh = \calh_C + \calh_T + \calh_D(t) ,
\edq
where $\calh_C=\sum_{k}\veps_{k}c_{k}^{\dag}c_{k}$
is the continuum of electron states with wavevector $k$ and band energy $\veps_{k}$,
$\calh_T=\sum_{k}(w_{k}d^{\dag}c_{k}+\text{h.c.})$
describes the tunneling hybridization between propagating electrons
and the localized fermion with coupling amplitude $w_{k}$,
and $\calh_D(t)=\veps_d(t) d^{\dag}d$ represents the impurity Hamiltonian
with a time-dependent energy level $\veps_d(t)=\veps_0 + V_{ac}\cos(\Omega t)$,
$\veps_0$ being the energy of the bare level. This model can be implemented,
e.g., using an electronic terminal coupled to a quantum dot acting as
an artificial impurity~\cite{gab06,feb07} which, in turn, is interacting with a nearby
capacitive gate with harmonic driving potential $V_{ac}\cos(\Omega t)$, where
$V_{ac}$ and $\Omega$ are the ac amplitude and frequency, respectively.
Our model is also relevant for fermionic gases of cold atoms~\cite{bra12}
in periodically driven optical lattices~\cite{sal09}.
For definiteness, we take a single reservoir in the spinless case but
the model can be straightforwardly generalized to account for multiple
leads and spinful electrons.

The Hamiltonian given by Eq.~\eqref{eq_H} conserves the number of particles
but not the total energy. We can write,
\beq \label{cons}
\frac{d\langle \calh\rangle}{dt}=W_C(t)+W_T(t)+W_D(t) +P(t)\,,
\edq
where the energy fluxes (energy per unit time) are $W_C(t)=i\langle [\calh,\calh_C]\rangle/\hbar$,
$W_T(t)=i\langle [\calh,\calh_T]\rangle/\hbar$ and $W_D(t)=i\langle [\calh,\calh_D]\rangle/\hbar$,
and fulfill $W_C(t)+W_T(t)+W_D(t)=0$. The term $P(t)= \langle\partial \calh_D/\partial t\rangle $ is the power developed by the ac forces.
Importantly, energy transport contains an additional term as compared to charge transport.
In the latter case, the current conservation condition
reads $I_C(t)+I_D(t)=0$ where the electronic currents (charge per unit time)
in the reservoir and the quantum level are given, respectively, by
$I_C(t)=ie\langle [\calh,\sum_{k}c_{k}^{\dag}c_{k}]\rangle/\hbar$
and $I_D(t)=ie\langle [\calh,d^\dagger d]\rangle/\hbar$.
There is no particle flux associated to the coupling Hamiltonian $\calh_T$
(although the currents must, of course, be calculated in the presence of $\calh_T$).
In stark contrast, the energy flux in the reservoir, $W_C(t)$, cannot be solely
inferred from that in the impurity, $W_D(t)$, but necessitates knowledge on how
energy is absorbed or desorbed in the contact region, $W_T(t)$.
This crucial fact introduces some ambiguity in the definition
of the concept of heat current, as shown below.

The different energy fluxes entering Eq.~\eqref{cons} can be computed 
in terms of the retarded 
$\calg^r(t,t') =- i\theta(t-t')\langle \{d(t),d^{\dag}(t')\} \rangle$ and lesser  $\calg^<(t,t') = i\langle d^{\dag}(t')d(t) \rangle$ Green functions. We find that
the energy flux entering the reservoir at time $t$ reads \cite{suppl}
\beq\label{wc}
W_C = - 2  \Re\int \frac{d\veps}{h} \Gamma(\veps) 
\left[ i \calg^r(t,\veps) f(\veps) \veps+ \calg^<(t,\veps) \Theta(\veps)\right]\,,
\edq
where  ${\cal G}(t,t^{\prime})= \int \frac{d \veps}{2 \pi} e^{-i \veps (t-t^{\prime})} {\cal G}(t,\veps)$ 
and $\Theta(\veps)= \int \frac{d \veps^{\prime}}{2 \pi}  \frac{\veps^{\prime}}{\veps-\veps^{\prime} -i 0^+}$.
In Eq.~\eqref{wc}, $f(x)=1/[1+e^{(x-\mu)/k_BT}]$ is the Fermi-Dirac distribution with
background temperature $T$, the chemical potential $\mu$, and
$\Gamma(\veps)=2 \pi \sum_k  |w_k|^2 \delta (\veps-\veps_k)$ is the resonance width
due to coupling to the continuous set of states. For definiteness,
we consider a model for the continuum with a flat density of states,
corresponding to a constant $\Gamma$.
We emphasize that  Eq.~\eqref{wc} is completely general and valid to all orders in $\Omega$ and $V_{ac}$. Moreover, it would be valid even in the presence of Coulomb
interactions acting on the spatially localized region.

Following the same procedure, we find for the impurity energy flux
the expression 
\beq \label{wd}
W_D(t)=-\veps_d(t)I_C(t)/e,
\edq
 where $I_C(t)=- 2 e \Re \int {d\veps}/h \Gamma(\veps) [ i \calg^r(t,\veps) f(\veps) + \calg^<(t,\veps) ]$
 is the charge current measured in the reservoir.
Equation~\eqref{wd} has a rather simple interpretation.
Let $n_d(t)$ be the expected value of the particle number at the localized site.
Then, its total energy rate of change is $d[\veps_d(t) n_d(t)]/dt$,
which consists of two terms, namely, the ac source power $P(t)
=n_d(t)d\veps_d/dt$
and the energy flux $W_D=\veps_d(t) dn_d/dt=-\veps_d(t)I_C(t)/e$,
since $I_D(t)\equiv e dn_d/dt=-I_C(t)$.

Finally, we determine the energy flux 
associated with the region that mixes 
continuous and localized states, $W_T=-W_C-W_D$.
It reads,
\beq\label{eq_wt}
W_T(t)= 2\Re\int \frac{d\veps}{h} \partial_t \calg^r(t,\veps) \Gamma f(\veps)\,,
\edq
with
$\calg^r(t,\veps)=\sum_n e^{-i n \Omega t} \calg(n,\veps)$. It is easy to verify that
Eq.~\eqref{eq_wt} is a purely ac contribution and vanishes
in the limit $\Omega\to 0$.  Thus,
for applied static fields or for time-averaged ac transport,
this special contribution to the system's energy flow is zero,
The quantity $W_T$ will be nonzero only for systems exhibiting a dynamical response.
In a quantum-dot setup, the tunnel barrier coupling the dot
and the contact lead would periodically store and release energy
in response to a nearby ac field, thereby the term {\em energy reactance}.

To gain further insight into the physical significance of $W_T$,
we now resort to the scattering-matrix formalism applied to quantum transport.
Equivalence between Green-function and scattering matrix approaches
has been proven in Ref.~\cite{arr06} for averaged time-dependent quantities.
But because $W_T$ precisely vanishes in the stationary limit, we now analyze
the full time-dependent energy flux by considering the energy current density operator
$\rho_E=\Psi^{\ast} \calh \Psi$, where
$\calh = -\hbar^2\nabla^2/2m + U(t,\vec{r}) $ is the
first-quantized version of Eq.~\eqref{eq_H} and $U$
is the full electronic potential which includes
externally applied time-dependent fields. Then, $\rho_E$
satisfies the continuity equation~\cite{mat}
\beq
\partial_t \rho_E + \nabla\cdot W_E = S_E,
\edq
where $W_E=(\hbar/4mi)[\Psi^{\ast}\calh\nabla\Psi - \nabla\Psi^{\ast}\calh\Psi + \text{h.c.}]$
is the symmetrized energy flux and $S_E=\Psi^{\ast}\partial_t U\Psi$
is the source term accounting for the explicit time dependence of $U$.
As is customary (see, e.g., Ref.~\cite{mos12}), we introduce the field operator
$\hPsi \sim \int d\veps\, e^{-i\veps t/\hbar}[ e^{+ikx} \ha(\veps) + e^{-ikx} \hb(\veps)]$
at the cross section $x$-position through which the flux is measured.
Then, the energy flux is expressed as
\begin{align}
W_E(t) &= \sum_{n,q} e^{-in\Omega t} \int d\veps \frac{\veps_{q}+\veps_{n+q} }{2h}
S^{F\ast}(\veps_q,\veps)S^{F}(\veps_{n+q},\veps)\nonumber\\
&\times[f(\veps_q)-f(\veps)]
\,,\label{eq_WE}
\end{align}
where the Floquet scattering matrix 
relates the lead outgoing flux operators $\hb$ to the incoming ones $\ha$ via
$\hb(\veps) = \sum_{n}S^{F}(\veps,\veps_n)\ha(\veps_n)$ and $\veps_n=\veps+n\hbar\Omega$.

Remarkably, if we now insert the generalized Fisher-Lee relation ~\cite{fis81,arr06}
$S^{F}(\veps_m,\veps_n) =  \delta_{m,n} - i \Gamma \calg( m-n,\veps_n) $
into Eq.~\eqref{eq_WE}  we find \cite{suppl}
\beq \label{we-wc}
W_E(t)= W_C(t) + \frac{1}{2}W_T(t)\,.
\edq
This relation states that in the presence of time-dependent fields
the energy fluxes entering the reservoir
predicted by scattering theory and the
Green function tunneling Hamiltonian approach
surprisingly differ by a term $\frac{1}{2}W_T$. 
Note that this departure occurs
for dynamical energy transport only. In the case of
time-dependent particle currents or time-averaged
energy fluxes the correspondence 
between the two theoretical frameworks 
is exact, i.e., $\overline{W_E}=\overline{W_C}$
with the notation $\overline{(\ldots)}=\int_0^\tau (\ldots) dt/\tau$, being $\tau=2 \pi/\Omega$.

What is the origin of the discrepancy in Eq.~\eqref{we-wc}?
Let us turn back to Fig.~\ref{fig1} and examine the role
of the contact region. While the scattering approach
considers propagating electrons with potential energy described
by the single function $U$,
the resonant level model considers partitions of the energy contributions as in Eq.~\eqref{eq_H},
similarly to Bardeen's picture of tunneling~\cite{bar61}.
Clearly, the mixing Hamiltonian $\calh_T$ contains creation and destruction operators associated to
degrees of freedom of electrons within the continuum as well as within the localized state. When separating the full setup into 
a reservoir and the driven localized part, it is then natural to split $\calh_T$ symmetrically, contributing equally to these two pieces. 
The point we would like to make here is that Eq. \eqref{we-wc}  shows that one should carefully
examine how heat fluxes are measured in a given setup
before attempting a detailed comparison with theory. 

A concomitant question is which portion of the energy flux
can be identified as heat. In stationary systems, where the heat transport is accompanied by the particle transport, the heat flux between the localized system and the reservoir is defined from the change
in the energy stored in the reservoir subtracting the convective term originated by the particle flow \cite{heat}. Such definition was also adopted for the dc component of the heat flux in time-dependent driven systems \cite{arr07}, obtaining the same description within the frameworks of the Green function 
and scattering matrix formalisms. However, there is an ambiguity in defining heat in the time domain. Specifically, Eq.~\eqref{eq_WE} suggests that the appropriate definition is 
\beq \label{q}
\dot{Q}(t)= W_E(t)- \mu I_C(t)/e=W_C(t)+ \frac{1}{2}W_T(t) - \mu I_C(t)/e,
\edq
while Eq. \eqref{cons} implies the heat flow definition $\dot{\tilde{Q}}= W_C(t)- \mu I_C(t)/e$. 

We resort to the basic principles of thermodynamics in order to argue that Eq.~\eqref{q} is the most meaningful definition of heat flux in the time-domain. Since the 
reservoir is a macroscopic system, a suitable interpretation of the different portions of its internal energy under slow variations of the driven localized part, would lead 
 to the definition of heat. We proceed along the lines of a textbook analysis \cite{balian},  identifying as  the reservoir the terms of the 
Hamiltonian $\calh$ containing operators $c_k^{\dagger}, c_k$ and as the driven system those depending on $d^{\dagger}, d$. The tunneling part
$\calh_T$ contains both, hence, it is natural to consider  the symmetric splitting $\calh_E=\calh_C+\frac{1}{2}\calh_T$ describing the the reservoir and $\calh_S(t)=\calh_D(t)+\frac{1}{2}\calh_T$ defining the driven system. We then evaluate the rate of change of the internal energy $ \dot{\langle \calh_E \rangle} = \dot{\langle \calh_C \rangle} - \frac{1}{2} \sum_{k} \left[ \veps_k - \veps_d(t) \right] \dot{n_k }$, with $n_k= \langle c^{\dagger}_k c_k \rangle $, which leads us to interpret
the quantity $\delta \langle \calh_T \rangle= -  \sum_{k} \left[ \veps_k - \veps_d(t) \right] \delta n_k $  as the chemical work due to particle flow through the contact. Hence,
in accordance to the first principle of thermodynamics, 
an appropriate definition for the heat exchange in the reservoir induced by slow variations of the driven system  is $ \delta Q= \delta \langle \calh_C \rangle + \delta \langle \calh_T \rangle /2 - \mu \delta N_C$, with $N_C= \sum_k n_k $, as suggested by  Eq.~\eqref{q}. In what follows we also show that this expression is also in agreement with the second law of thermodynamics, while this is not the case of the alternative definition $\dot{\tilde{Q}}$.

We focus on the slow driving regime and consider, for simplicity, zero temperature ($T=0$). 
Then, an exact analysis can be performed by means of an expansion in powers of
$\Omega$ for the Green functions (or equivalently of the scattering matrix) \cite{mobu2004}:
\beq
\calg^{r}(t,\veps) = \calg_f^{r}(t,\veps) + \frac{i\hbar}{2} \partial_t\partial_{\veps} \calg_{f}^{r}(t,\veps)+\ldots
\edq
$\calg_f^{r} (t,\veps)=[\veps-\veps_d(t)+i\Gamma /2 ]^{-1}$ is the \textit{frozen} Green function
describing the regime in which the electron instantaneously
adjusts its potential to the ac field.  Considering the expansion of $\calg$ up to ${\cal O}(\Omega)$ 
yield heat fluxes exact up to ${\cal O}(\Omega^2)$ \cite{suppl}. We find
$\dot{Q}(t)= \dot{Q}^{(1)}(t)+\dot{Q}^{(2)}(t)$, where the first and second order terms in $\Omega$ are, respectively, 
\begin{eqnarray}
\dot{Q}^{(1)}(t)& = & \int\frac{d\veps}{h}(\mu -\veps)\frac{\partial f}{\partial \veps} \rho^f (t,\veps)\frac{ d {\veps_d}}{dt}, \label{q1}\\
\dot{Q}^{(2)}(t) & = &  -\frac{1}{2}\int\frac{d\veps}{h}\frac{\partial f}{\partial \veps}\{(\mu -\veps)\frac{d}{dt} \left[{ [ {\rho^f}}(t,\veps) ]^2 \frac{d {\veps_d}}{dt}\right]
\nonumber \\
& & 
+ \left[ {\rho^f}(t,\veps) \frac{d \veps_d}{dt} \right]^2 \}. \label{q2}
\end{eqnarray}
Here $\rho^f(t,\veps)= -2 \mbox{Im}[\calg_f^r(t,\veps)]= | {\cal G}_f^r(t,\veps)|^2 \Gamma = -i \partial_{\veps} S_f S_{f}^{\ast}$ is the local density of states and
$S_f(t,\veps)$ 
the frozen scattering matrix, i.e,
the stationary scattering matrix with time-dependent parameters.

Both the first-order term $\dot{Q}^{(1)}(t)$ and the first term of $\dot{Q}^{(2)}(t)$
vanish at $T=0$ since $-\partial_\veps f = \delta(\veps-\mu)$.
The component $\dot{Q}^{(2)}(t)$, which is second order in $\Omega$, 
 represents the leading-order to the dissipated power in the reservoir.  At $T=0$, Eq. \eqref{q2} reduces to $\dot{Q}^{(2)}(t)= [ {\rho^f}(t,\mu) \frac{d \veps_d}{dt} ]^2/2$.
Evaluating the charge current up to the first order in $\Omega$, we find, $ I_{C}^{(1)}(t) = -(e/h) \rho^f (t,\mu)\frac{d \veps_d}{dt}$,
which implies  
\beq
\dot{Q}^{(2)}(t)= R_q [I_C^{(1)}(t)]^2,
\edq
with $R_q=h/2e^2$ the relaxation resistance quantum~\cite{but93a,but93b}.
Since $R_q$ is a manifestly positive quantity at all times,
the heat flux given by Eq.~\eqref{q} represents the heat dissipated into the cold reservoir
when the system is coupled to the ac driving force. Therefore,
Eq.~\eqref{q} agrees with the second law of thermodynamics.

\begin{figure}[!h]
\begin{center}
\includegraphics[width=0.45\textwidth]{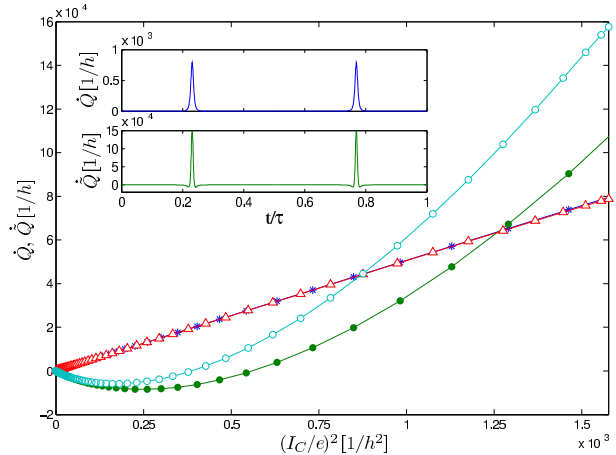}
\caption{(Color online). Heat fluxes $\dot{Q}(t)$ (stars and triangles)
and $\dot{\tilde{Q}}(t)$ (solid and open circles)  
as a function of the charge current $I_C(t)^2$ within the slow driving regime
for two different amplitudes $V_{ac}=10,12$, respectively.
Clearly, only the heat $\dot{Q}(t)$ satisfies
$\dot{Q}(t)/I_C(t)^2=R$ with $R$ a constant independent of time.
Parameters: $\mu=0$, $\veps_0=-1.2$, $T=0$ and $\hbar \Omega=10^{-3}$.
Energies are expressed in units of $\Gamma$. Inset: $\dot{Q}(t)$  (dashed lines with a vertical offset) and $\dot{\tilde{Q}}(t)$ (solid lines) as a function of time.}
\label{fig2}
\end{center} 
\end{figure}

We reinforce our conclusion by comparing with the heat rate of change given by
$\dot{\tilde{Q}}$. Thus, we evaluate $W_T$ up to second order in $\Omega$:
\begin{eqnarray}
 W_T^{(1)} (t)& = &
          2 \int \frac{d \veps}{h} \frac{  \partial   f  }{\partial \veps }  
   \left[  \rho^f (t, \veps) \left(\veps - \veps_d(t) \right)\frac{ d \veps_d }{dt} \right]  , \nonumber\\   
  W_T^{(2)} (t)& = & - \int \frac{d \veps}{h} \frac{  \partial   f  }{\partial \veps }  
   \frac{d}{dt} \left[ [\rho^f (t,\veps)]^2 \left(\veps - \veps_d(t) \right) \frac{ d \veps_d }{dt} \right] . \nonumber \\
  \end{eqnarray} 
Within the weak driving regime,  $\dot{\tilde{Q}}(t) = \dot{Q}(t) -[ W_T^{(1)} (t)+ W_T^{(2)} (t)]/2$, which at $T=0$ contains contributions $\propto \Omega$ and $\propto \Omega^2$.
Defining the resistance $\tilde{R}(t)$ from the relation $\dot{\tilde{Q}}(t)=[I_C^{(1)}(t)]^2 \tilde{R}(t)$, we find that $\tilde{R}(t)$ is non-universal and depends on time.
In fact, it is not even positive definite and then $\dot{\tilde{Q}}(t)$ cannot
be interpreted as a dissipated heat. We illustrate in Fig.~\ref{fig2} the behavior of the two expressions of the heat flux for different amplitudes of the driving potential $V_{ac}$ for a reservoir at $T=0$ and small driving frequencies. The inset shows that, as a function of time, $\dot{Q}(t)$ is always positive whereas
$\dot{\tilde{Q}}(t)$ may attain negative values. The main panel shows $\dot{Q}(t)$ and $\dot{\tilde{Q}}(t)$ as a function of $I_C(t)^2$ within the slow driving regime. In the first case,
we observe a linear function with the universal slope $R_q$. In contrast, in the second case we observe a non-universal behavior, including negative values of $\tilde{R}(t)$. 
The two definitions of heat, however, lead to the \textit{same} result when averaged in time, $\overline{\dot{Q}}=\overline{\dot{\tilde{Q}}}= \overline{P}$ and, therefore,
only a pure \textit{dynamical} measurement would be able to distinguish both.

In conclusion, we have discussed the dynamical heat generation in a resonant level system due to coupling to an external time-dependent potential and highlighted the important role played by the energy associated with the coupling region. The latter is unique to dynamical energy transport.
By recourse to an adiabatic expansion valid for the slow-driving regime,
we have found that an appropriate expression of the dynamical heat flux that agrees with the
fundamental principles of thermodynamics requires to take into account the work associated
to particles flowing through the tunneling region. Importantly, the time-dependent
flux is \textit{instantaneously} given by a Joule law with universal resistance.
Our results are relevant for recent developments in the energetics of atomic systems and nanostructures.

We acknowledge M. B\"uttiker and R. L\'opez for numerous discussions.
This work was supported in part by MINECO Grants Nos.~FIS2011-23526
and CSD2007-00042 (CPAN) and the NSF under Grant No.~NSF~PHY11-25915, as well by UBACyT, CONICET and MINCyT, Argentina.

\setcounter{figure}{0}
\setcounter{equation}{0}
\renewcommand{\thefigure}{S\arabic{figure}}  
\renewcommand{\theequation}{S\arabic{equation}} 
\renewcommand{\figurename}{Figure} 
\renewcommand{\thefootnote}{S\arabic{reference}} 

\pagebreak
\onecolumngrid
\pagebreak

\section*{Dynamical energy transfer in ac driven quantum systems: Supplementary Information}

This supporting document describes further details of the derivation
of the time-dependent energy fluxes, a comparison with the scattering
matrix approach and the heat flow for the slow driving case.

\subsection{Time-dependent energy fluxes}

We consider a simple model of a harmonically driven resonant level, which is coupled to a single reservoir consisting of a fermionic band of continuous state with a flat density,
described by the Hamiltonian of Eq.~(1) 

In order to define the energy fluxes entering each part of the system, we analyze the evolution in time of the total energy:
\be
\frac{d\langle \calh\rangle}{dt}=\frac{d\langle \calh_C\rangle}{dt}+\frac{d\langle \calh_T\rangle}{dt}+\frac{d\langle \calh_D\rangle}{dt},
\ee  

The energy current flowing into the reservoir drives its energy variation, and as a consequence we define the energy flux entering the reservoir $W_C$  as
\be
\label{reservoir}
W_C\equiv\frac{d\langle \calh_C\rangle}{dt}=\frac{i}{\hbar}\langle[\calh,\calh_C]\rangle.
\ee
The same occurs for the energy variation stored in the coupling region. Thus, the energy flow reads
\be
\label{contact}
W_T\equiv\frac{d\langle \calh_T\rangle}{dt}=\frac{i}{\hbar}\langle[\calh,\calh_T]\rangle.
\ee
On the other hand, the rate of the energy in the driven system involves the energy flux entering it and the power developed by the time dependent voltage applied to the resonant level,
\be
\frac{d\langle \calh_D\rangle}{dt}=\frac{i}{\hbar}\langle[\calh,\calh_D]\rangle+\langle\frac{\partial\calh_D}{\partial t}\rangle.
\ee
From this expression, we identify the term involving the conmutator of operators as the energy flux entering the level $W_D\equiv\frac{i}{\hbar}\langle[\calh,\calh_D]\rangle$, and the power applied by the external field ${P}(t)=\langle\frac{\partial\calh_D}{\partial t}\rangle$.

Therefore, we can write
\be
\tag{2}
\label{fulfill}
\frac{d\langle \calh\rangle}{dt}=W_C(t)+W_T(t)+W_D(t)+P(t).
\ee

The energy of the full system is not conserved due to the presence of a power source, 
\be
\label{h}
\frac{d\langle \calh\rangle}{dt}=\langle\frac{\partial\calh}{\partial t}\rangle=\langle\frac{\partial\calh_D}{\partial t}\rangle=P(t),
\ee
and then combining Eqs.(\ref{fulfill}) and (\ref{h}), the energy fluxes fulfill the condition
$W_C(t)+W_T(t)+W_D(t)=0$.

Following definition (\ref{reservoir}), we start computing the energy flux entering the reservoir as a function of time,
\ba
W_C & = & \frac{i}{\hbar} \langle[\calh, \calh_C]\rangle =\frac{i}{\hbar}  \langle[\calh_T, \calh_C]\rangle \nonumber \\
&=& -\frac{i}{\hbar} \sum_{k}  \veps_k [w_{k}\langle c^{\dagger}_k(t) d(t)\rangle-{w_{k}}^{*}\langle d^{\dagger}(t) c_k(t) \rangle].
\ea 
Taking quantum-mechanical averages and
using the definition ${\cal G}^<_{k}(t,t^{\prime})=i \langle c^{\dagger}_k(t^{\prime}) d(t) \rangle $,
the variation in time of the energy stored in the reservoir can be written as follows:
\be
W_C= -\frac{2}{\hbar} \sum_{k}  \veps_k \mbox{Re}\{w_{k}{\cal G}^<_{k}(t,t)\}.
\ee
From Dyson equation and Langreth rules (see Refs.~\cite{lili1,lili2}), the above expression can be expressed as follows:
\ba \label{eq1}
W_C & = & -\frac{2}{\hbar}\sum_{k} \mbox{Re} \{ |w_{k}|^2 \,\veps_k
\int d t_1 [{\cal G}^r(t,t_1) g_k^<(t_1,t) + {\cal G}^<(t,t_1) g_k^a(t_1,t)] \}, 
\ea
being
\ba
{\cal G}^r(t,t_1)& = &-i\theta(t-t_1)\langle\{d(t),d^\dagger (t_1)\}\rangle,\nonumber \\
{\cal G}^<(t,t_1)& = &i\langle d^\dagger (t_1)d(t)\rangle,
\ea
and
\ba
g_k^<(t_1,t) & = & i \int \frac{d \veps}{ 2\pi} f(\veps) \gamma_k(\veps) e^{-i \veps(t_1-t)/\hbar}, \nonumber \\
g^a_k(t_1,t) & = & \int \frac{d \veps}{ 2\pi} \int\frac{d \veps^{\prime}}{ 2\pi} 
\frac{\gamma_k(\veps^{\prime})}{\veps-\veps^{\prime} - i 0^{+}} e^{-i \veps(t_1-t)/\hbar}, \nonumber \\
\gamma_k(\veps) &= & 2 \pi \delta(\veps - \veps_k).
\ea
Substituting in Eq. (\ref{eq1}):
\be \label{eq2}
W_C   =  -\frac{2}{\hbar}\mbox{Re}\{ \int dt_1 \int \frac{d\veps}{2\pi} \,e^{-i \veps (t_1 -t)/\hbar}\left[
i {\cal G}^r(t,t_1) {\Gamma} f(\veps) \veps +\int  \frac{d\veps^{\prime}}{2\pi}
{\cal G}^<(t,t_1) {\Gamma} \frac{\veps^{\prime} }{\veps - \veps^{\prime} - i \eta}\right] \},
\ee
where ${\Gamma}=
\sum_k |w_{k}|^2 \gamma_k(\veps)$ is the hybridization width due to coupling to the reservoir, and $f(\veps)=1/[1+e^{(\veps-\mu)/{k_BT}}]$ the Fermi- Dirac distribution. We also assume a reservoir with a wide-band density of states, such that $\Gamma$ is approximately independent of $\veps$.

Introducing the Fourier representation for the Green function
\be
{\cal G}(t,t_1) = \int \frac{d\veps}{2\pi}
e^{- i \veps (t-t_1)/\hbar }{\cal G}(t,\veps),
\ee
we obtain Eq.~(3) of the main text
\be
\tag{3}
\label{la3}
W_C=-2\mbox{Re}\{\int\frac{d\veps}{h}\Gamma\left[i{\cal G}^r(t,\veps)f(\veps)\veps+{\cal G}^<(t,\veps)\Theta_{\veps}(\veps)\right]\},
\ee
where
\be
\Theta(\veps)=\int  \frac{d\veps^{\prime}}{2\pi}\frac{\veps^{\prime} }{\veps - \veps^{\prime} - i 0^{+}}.
\ee

We now consider the Keldysh equation for the lesser Green function:
\be \label{dyles}
{\cal G}^<(t,\veps)= \int dt_1 \int dt_2 \int dt_3 {\cal G}^r(t,t_2) {\Sigma}^<(t_2-t_3) {\cal G}^a(t_3,t_1)e^{i\veps(t-t_1)/\hbar},
\ee
with ${\Sigma}^<(\veps) = i f(\veps) {\Gamma}$, and taking into account the fact that the forcing is periodic in time we introduce the following Floquet-Fourier representation for the Green function \cite{lili-moskalets}:
\be \label{florep}
{\cal G}^r(t,\veps) = \sum_n  e^{-i n \Omega t}{\cal G}(n,\veps),
\ee
$\Omega$ being the fundamental driving frequency.
Substituting Eqs.~(\ref{dyles}) and (\ref{florep}) in Eq. (\ref{la3}) we obtain:
\ba
W_C(t) & = &  -2 \mbox{Re} 
\{ \sum_l e^{-i l \Omega t} \int \frac{d\veps}{h}  
{\Gamma}\Big[ i {\cal G}^r(l,\veps)f(\veps)  \veps \nonumber \\
& & + i 
{\Gamma}\sum_n {\cal G}^r(l+n,\veps-n\hbar\Omega)  
{{\cal G}^r}^{*}(n,\veps-n\hbar\Omega)f(\veps - n \hbar\Omega)\Theta(\veps)  
\Big]\}.
\ea

Thus, after working with the above expression we finally find
\ba \label{ef}
W_C(t)  & = & -\sum_l e^{-i l \Omega t} \int \frac{d\veps}{h}  
\{   
i {{\cal G}^r}^{*}(-l,\veps) {\Gamma}
 [(\veps -l \hbar\Omega) f(\veps-l \hbar\Omega)  -
\veps f(\veps)  ] - \nonumber \\
& & \sum_{n} 
[(\veps + \frac{l\hbar\Omega}{2}) f(\veps-n\hbar\Omega)- \veps f(\veps)]
  {\cal G}^r(l+n,\veps-n\hbar\Omega){\Gamma}^2
{{\cal G}^r}^{*}(n,\veps-n\hbar\Omega) \}.
\ea

Following a similar procedure, we can compute the other fluxes entering Eq.~(\ref{fulfill}). The energy current flowing through the contact between the reservoir and the impurity reads
\ba
W_T(t) & = &\sum_{k} \left[w_{k}\frac{d\left< d^{\dagger}(t) c_k(t)\right>}{dt} + h.c\right]=2\sum_{k}{\mbox{Im}\left\{w_{k}\frac{d{\cal G}_{k}^{<}(t,t)}{dt}\right\}}\nonumber\\
& = & \int\frac{d\veps}{h}\Omega f(\veps){\Gamma}\sum_l l\left[2\mbox{Im}\{e^{-i l\Omega t}{\cal G}^r(l,\veps)\}+\sum_n{\Gamma}\mbox{Re}\{e^{-i l\Omega t}{\cal{G}}^r(l+n,\veps){{\cal G}^r}^{*}(n,\veps)\}\right].
\ea

It is easy to prove that the last term $\sum_{l,n} l \,\mbox{Re}\{e^{-i l\Omega t}{\cal{G}}^r(l+n,\veps){{\cal{G}}^r}^{*}(n,\veps)\}$ vanishes,
\ba
\sum_{l>0}\sum_n\left(l\,\mbox{Re}\{e^{-i l\Omega t}{\cal{G}}^r(l+n,\veps){{\cal{G}}^r}^{*}(n,\veps)\}-l\,\mbox{Re}\{e^{i l\Omega t}{\cal{G}}^r(-l+n,\veps){{\cal{G}}^r}^{*}(n,\veps)\}\right)=\nonumber
\ea
\ba
\sum_{l>0}\sum_n\,l\,\left(\mbox{Re}\{e^{-i l\Omega t}{\cal{G}}^r(l+n,\veps){{\cal{G}}^r}^{*}(n,\veps)\}-\mbox{Re}\{e^{i l\Omega t}{{\cal{G}}^r}^{*}(l+n,\veps){{\cal{G}}^r}(n,\veps)\}\right)=0.\nonumber
\ea

As a consequence, the variation of the energy stored in the contact is 
\ba
\label{diff}
W_T(t) 
& = & \int\frac{d\veps}{h}\Omega f(\veps)\sum_l l\,2\mbox{Im}\{e^{-i l\Omega t}{\cal{G}}^r(l,\veps){\Gamma}\}.
\ea

Combining this expression with Eq.~(\ref{florep}), we find
\be
\tag{5}
W_T(t)=2 \mbox{Re}\{\int \frac{d\veps}{h}\partial_t{\cal{G}}^r(t,\veps){\Gamma}f(\veps) \}.
\ee

Finally, for the energy flux entering the impurity we find
\be
W_D = \frac{i}{\hbar}\langle[\calh ,\calh_D]\rangle =\veps_d(t)\frac{dn_d(t)}{dt}, 
\ee
where $n_d(t)=\langle d^{\dagger}(t)d(t)\rangle$ is the number of particles present in the resonant level, which is related to the charge current measured in the reservoir $I_C$ due to the conservation of the charge
\be
e\frac{dn_d(t)}{dt}=-I_C(t). 
\ee

Thus, 
\be
\tag{4}
W_D=-\veps_d(t)I_C(t)/e.
\ee

We apply the same procedure as before to obtain an expression for the charge current flowing through the contact between the reservoir and the system, defined as
\ba
I_{C}(t)& = &\frac{ie}{\hbar}\langle[\calh,\sum_{k}c^{\dagger}_{k}c_k]\rangle=-\frac{2e}{\hbar}\mbox{Re}\{w_k {{\cal G}_k}^<(t,t)\}\nonumber\\
& = & \frac{e}{h}\mbox{Im}\{\int d\veps \Gamma\left[2{\cal G}^r(t,\veps)f(\veps)+{\cal G}^<(t,\veps)\right]\}.
\ea 
Using Eq.~(\ref{florep}), this quantity can be written as
\ba
\label{charge}
I_C(t)  & = & -\frac{e}{h}\sum_l e^{-i l \Omega t} \int {d\veps}  
\{   
i {{\cal G}^r}^{*}(-l,\veps) {\Gamma}
 [ f(\veps-l\hbar \Omega)  -
f(\veps)  ] - \nonumber \\
& & \sum_{n} 
[ f(\veps-n\hbar\Omega)-  f(\veps)]
  {\cal G}^r(l+n,\veps-n\hbar\Omega){\Gamma}^2
{{\cal G}^r}^{*}(n,\veps-n\hbar\Omega) \}.
\ea

\subsection{Comparison with the scattering matrix approach}

Previous works  \cite{lili-moskalets,trans2} have proven the existence of a simple relation between the scattering matrix elements and the Fourier coefficients for the Green function. For periodically driven systems, the Fisher-Lee formula for stationary systems has been generalized as
\be
\label{sm-green}
S^{F}(\veps_m,\veps_n)=\delta_{m,n}-i\Gamma {\cal G}^{r}(\veps_{m-n},\veps_n),
\ee
where $S^{F}(\veps,\veps_n)$ are the Floquet scattering matrix elements, and $\veps_n=\veps+n\hbar\Omega$.
The above relation leads to expressions for the time-averaged charge and energy currents that are equivalent in both formalisms. 
The aim of the present section is to show in detail the derivation of Eq.~(8) of the main text.

To start, according to Ref. \cite{notes} the time-dependent energy flow within the scattering matrix formalism reads:
\be
\tag{7}
W_E(t) = \sum_{n,q} e^{-i n \Omega t} \int d\veps \frac{\veps_q+ \veps_{n+q}}{2h} {S^F}^*(\veps_q,\veps) S^F(\veps_{n+q},\veps)[f(\veps)-f(\veps_q)],
\ee
and taking into account the relation (\ref{sm-green}), the above expression
can be written in terms of Green functions:
\ba
\label{scatt}
W_E(t) & = & -\sum_l e^{-i l \Omega t} \int \frac{d\veps}{h}  
\{   
i {{\cal G}^r}^{*}(-l,\veps){\Gamma}  
 \,(\veps -\frac{l}{2}\hbar \Omega)[ f(\veps-l \hbar\Omega)  - f(\veps)  ] - \nonumber \\
& & \sum_{n} 
(\veps + \frac{l\hbar\Omega}{2}) [f(\veps-n\hbar\Omega)- f(\veps)]
{\cal G}^r(l+n,\veps-n\hbar\Omega) {\Gamma}^2
{{\cal G}^r}^{*}(n,\veps-n\hbar\Omega)\}.
\ea
Comparing with the expression (\ref{ef}) it can be seen that
\ba
W_E(t)-W_C(t) & = &  \sum_l e^{-i l \Omega t} \int \frac{d\veps}{h}  
\{   
i {{\cal G}^r}^{*}(-l,\veps) {\Gamma}  
 (-\frac{l}{2} \hbar\Omega) [ f(\veps-l\hbar \Omega)  + f(\veps)  ] - \nonumber \\
& & \sum_{n} 
\frac{l\hbar\Omega}{2} f(\veps)
 {\cal G}^{r}(l+n,\veps-n\hbar\Omega){\Gamma}^{2} 
{{\cal G}^{r}}^{*}(n,\veps-n\hbar\Omega)  \} \nonumber \\
& = &  - \sum_l e^{-i l \Omega t} \int \frac{d\veps}{h}  
   \frac{l\hbar\Omega}{2} f(\veps){\Gamma}\{
i {\cal G}^{r}(l,\veps)+
i {{\cal G}^{r}}^{*}(-l,\veps) \}\nonumber \\
&=& \int \frac{d\veps}{h} f(\veps)\hbar\Omega\sum_l l\,\mbox{Im}\{ e^{-i l \Omega t}{\cal G}^{r}(l,\veps){\Gamma} \}.
\ea

From Eq.~(\ref{diff}) we can see that this difference is related to the energy stored in the contact,
\be
\tag{8}
W_E(t)-W_C(t)=\frac{1}{2}W_T(t).
\ee

This is a surprising result because
in addition to presenting a discrepancy between the energy fluxes
predicted by scattering theory and the tunneling Hamiltonian model
it also states that the difference is $\frac{1}{2}W_T(t)$ related
to the energy flowing through the contact. 
This result does not contradict the exact correspondence within the stationary limit, i.e., $\overline{W_E}=\overline{W_C}$, because $W_T$ vanishes when it is averaged over time.

\subsection{Heat flow for slow driving}

For slow driving, an exact solution of the Dyson equation up to ${\cal O}(\Omega)$ can be obtained by expanding the Green functions in powers of $\Omega$ \cite{mobu2004si}

\be
\tag{10}
{\cal G}^{r}(t,\veps)={\cal G}^{r}_{f}(t,\veps)+\frac{i\hbar}{2}\partial_t \partial_{\veps}{\cal G}^{r}_{f}(t,\veps),
\ee
where ${\cal G}^{r}_{f}=[\veps-\veps_d(t)+i\Gamma/2]^{-1}$ is the frozen Green function, and its derivatives are
\ba
\partial_{\veps}{\cal G}^{r}_{f}(t,\veps) & = & -{{\cal G}^{r}_{f}(t,\veps)}^{2}\nonumber\\
\partial_t {\cal G}^{r}_{f}(t,\veps) & = &-\partial_{\veps}{\cal G}^{r}_{f}(t,\veps)\frac{\veps_d(t)}{dt}.
\ea

In this section we present the expressions for both definitions of the heat flow
\ba
\label{heatdefinition}
\dot{Q}(t) & = & W_E(t)-\mu I_C(t)/e\\
\dot{\tilde{Q}}(t) & = & W_C(t)-\mu I_C(t)/e=\dot{Q}(t) -\frac{1}{2}W_T(t)\nonumber
\ea
within that approximation.

Thus, we just need to compute the energy flux entering the reservoir $W_E$ and the charge current $I_C$ at low frequencies. We can express those quantities as a sum of linear and quadratic terms in the driving frequency $\Omega$, respectively
\ba
W_E(t)=W_E^{(1)}(t)+W_E^{(2)}(t)\nonumber\\
I_C(t)=I_C^{(1)}(t)+I_C^{(2)}(t).
\ea
The zero order term for those quantities vanishes and is not present in the above expansion.
  
Next, we expand $f(\veps+n\hbar\Omega)\sim f(\veps)+({\partial f}/{\partial\veps})n\hbar\Omega+({\partial^{2}f}/{\partial\veps^{2}})(n\hbar\Omega)^{2}/2$ and
\be
{\cal G}(n,\veps)\sim{\cal G}^{(0)}(n,\veps)+\hbar\Omega{\cal G}^{(1)}(n,\veps)
\ee
with
\ba
{\cal G}^{(0)}(n,\veps)& = & \int_{0}^{\tau}\frac{dt}{\tau}{\cal G}^{r}_{f}(t,\veps)e^{in\Omega t}\nonumber\\
\Omega {\cal G}^{(1)}(n,\veps)& = & \int_{0}^{\tau}\frac{dt}{\tau}\frac{i}{2}\partial_t \partial_{\veps}{\cal G}^{r}_{f}(t,\veps)e^{in\Omega t},
\ea
where $\tau=2\pi/\Omega$.
Substituting these expansions into Eqs.~(\ref{charge}) and~(\ref{scatt}), and performing an inverse Fourier transformation, we can express the energy and charge currents in terms of the frozen Green function:
\ba
{W_E}^{(1)}(t) & = & -\int \frac{d\veps}{h}\frac{\partial f}{\partial\veps}\,\veps \, \rho^f(t,\veps)\frac{d\veps_d}{dt}\nonumber\\
{W_E}^{(2)}(t) & = &\frac{1}{2}\int\frac{d\veps}{h}\frac{\partial f}{\partial\veps}\{\veps\frac{d}{dt}\Big([\rho^{f}(t,\veps)]^2\frac{d\veps_d}{dt}\Big)-\Big({\rho^f}(t,\veps)\frac{d\veps_d}{dt}\Big)^2\}, 
\ea
and
\ba
{I_C}^{(1)}(t) & = & -\frac{e}{h}\int{d\veps}\frac{\partial f}{\partial\veps}\, \rho^f(t,\veps)\frac{d\veps_d}{dt}\nonumber\\
{I_C}^{(2)}(t) & = &\frac{e}{2h}\int{d\veps}\frac{\partial f}{\partial\veps}\,\frac{d}{dt}\Big([\rho^{f}(t,\veps)]^2\frac{d\veps_d}{dt}\Big),
\ea
where we have defined the \textit{frozen} density of state $\rho^f(t,\veps)=-2\mbox{Im}\{{\cal G}^{r}_{f}(t,\veps)\}=\Gamma\mid{\cal G}^{r}_{f}(t,\veps)\mid^{2}$.

Now, we can expand the heat flow $\dot{Q}(t)$ in the same fashion
and obtain from Eq.~(\ref{heatdefinition})
\be
\tag{11}
{\dot{Q}}^{(1)}(t) = \int \frac{d\veps}{h}\frac{\partial f}{\partial\veps}\,(\mu-\veps) \, \rho^f(t,\veps)\frac{d\veps_d}{dt}
\ee
\be
\tag{12}
{\dot{Q}}^{(2)}(t)  = -\frac{1}{2}\int\frac{d\veps}{h}\frac{\partial f}{\partial\veps}\{(\mu-\veps)\frac{d}{dt}\Big([\rho^{f}(t,\veps)]^2\frac{d\veps_d}{dt}\Big)+\Big({\rho^f}(t,\veps)\frac{d\veps_d}{dt}\Big)^2\}. 
\ee

At temperature $T=0$, since $\partial_{\veps}f\sim\delta(\veps-\mu)$, both the first order term $\dot{Q}^{(1)}(t)$ and the first term of $\dot{Q}^{(2)}(t)$ vanish. In contrast, the last term of the second order component of the heat is equal to the power developed by the external voltage $P(t)$, since
\be
P(t)=\langle\frac{\partial\calh_D}{\partial t}\rangle=\frac{d \veps_d}{dt}\langle n_d(t)\rangle,
\ee 
which in the low frequency approximation reads
\be
P(t)=-\frac{d \veps_d}{dt}\frac{1}{2}\int\frac{d\veps}{h}\frac{\partial f}{\partial\veps}\,\Big([\rho^{f}(t,\veps)]^2\frac{d\veps_d}{dt}\Big).
\ee

To finalize, in order to compute the alternative definition of heat $\dot{\tilde{Q}}(t) $ within this approximation, we have to evaluate $W_T$ up to second order in $\Omega$. Following the same procedure from Eq. (\ref{diff}) we find:
\ba
 W_T^{(1)}(t) & = & - 2\int \frac{d \veps}{h} f(\veps) 
\mbox{Re}\{ \frac{  \partial  {\cal{G}}^r_f (t, \veps) }{\partial \veps}  \}{\Gamma}\frac{d\veps_d}{dt}\nonumber \\
     &=&  2 \int \frac{d \veps}{h} \frac{  \partial   f(\veps) }{\partial \veps }  
  \left[{\rho}^f (t, \veps) \left(\veps - \veps_d(t) \right)\frac{ d \veps_d}{dt}  \right]\nonumber\\ 
 W_T^{(2)}(t) & = & 2\int \frac{d \veps}{h}f(\veps)\frac{d}{dt}\mbox{Im}\{{\cal{G}}_f ^r(t, \veps) \frac{  \partial  {\cal{G}}_f^r (t, \veps) }{\partial t} {\Gamma}\}\nonumber\\
& = &-\int \frac{d \veps}{h} \frac{  \partial   f(\veps) }{\partial \veps }  
   \frac{d}{dt}\Big([\rho^{f}(t,\veps)]^2 \left(\veps - \veps_d(t) \right) \frac{ d \veps_d}{dt}\Big),
\ea
where we have used the relation between real and imaginary parts of the Green function
\be
\mbox{Re}\{{\cal G}_f^{r}(t,\veps)\}=-2\mbox{Im}\{{\cal G}_f^{r}(t,\veps)\}(\veps-\veps_d(t))=\rho^{f}(t,\veps)(\veps-\veps_d(t)).
\ee

\end{document}